\documentclass[aps,prb,showpacs,twocolumn,superscriptaddress,11]{revtex4-2} 
\pdfpageattr {/Group << /S /Transparency /I true /CS /DeviceRGB>>}

\usepackage{graphicx}		
\usepackage{amsmath}
\usepackage{physics}
\usepackage{bm}
\usepackage{hyperref}
\hypersetup{%
	pdftitle={Feshbach resonance of heavy exciton-polaritons},%
	pdfauthor={Brian Vermilyea and Michael M. Fogler},%
	pdfpagemode={UseNone},%
	pdfstartview={FitH},%
	breaklinks=true,%
	citecolor=blue,%
	colorlinks=true,%
	linkcolor=blue,%
	urlcolor=blue
}
\usepackage[T1]{fontenc}
\usepackage{fourier}
\usepackage{baskervald}
\DeclareMathAlphabet{\mathcal}{OMS}{cmsy}{m}{n}

\begin{document}
\author{B. Vermilyea and M.M. Fogler \\
\small{\textit{Department of Physics, University of California San Diego, 9500 Gilman Drive, La Jolla, California, 92093, USA}}}
\title{Feshbach resonance of heavy exciton-polaritons}
\begin{abstract}
	We study interactions between polaritons formed by hybridization of excitons 
	in a two-dimensional (2D) semiconductor with surface optical phonons or plasmons.
	These quasiparticles have a high effective mass and
	can bind into bipolaritons near a Feshbach-like scattering resonance.
	We analyze the phase diagram of a many-body condensate of heavy polaritons and bipolaritons
	and calculate their absorption and luminescence spectra, which can be measured experimentally.
\end{abstract}
\maketitle

\section{Introduction}
\label{sec:introduction}

Exciton-polaritons, formed by coherent 
coupling of excitons with photons in semiconductor microcavities, 
have been a subject of active research \cite{Hopfield1958, Kavokin2011}.
These hybrid light-matter quasiparticles are of interest due to their potential device applications \cite{Sanvitto2016}
and solid-state realizations of
condensation and superfluidity \cite{Deng2010, Carusotto2013}.
Many phenomena unique to exciton-polaritons stem from
strong interactions induced by their excitonic component.
Recent pump-probe optical experiments revealed that exciton-polaritons
can exhibit a Feshbach scattering resonance mediated by the biexciton state 
\cite{Wouters2007, Carusotto2010, Takemura2014,
Takemura2017, NavadehToupchi2019, BastarracheaMagnani2019},
which in some sense allows one to control the strength and sign of polariton interactions,
akin to the manipulation performed in experiments with atomic gases.

The origin of the exciton-polariton Feshbach resonance is associated
with formation of bipolaritons, i.e., bound states of polaritons. 
One might expect that bipolaritons are realized when the Rabi splitting (Sec.~\ref{sec:single-particle})
exceeds the biexciton binding energy. 
However, due to the strong dispersion of the photon, exciton-polaritons have a very small effective mass
and the bipolariton binding energy is exponentially suppressed in 2D.
Thus, bipolariton states are not observed \cite{Borri2000},
although radiative corrections can significantly modify the biexciton dispersion 
\cite{Ivanov1993, Ivanov1995, Ivanov1998, Ivanov2004}.
Note that while such radiatively renormalized biexciton states are often called bipolaritons in the literature,
in this paper we use the term bipolariton to mean a true bound state of two polaritons
with energy below the two-particle continuum.

\begin{figure}[t]
\begin{center}
\includegraphics[width=3.2in]{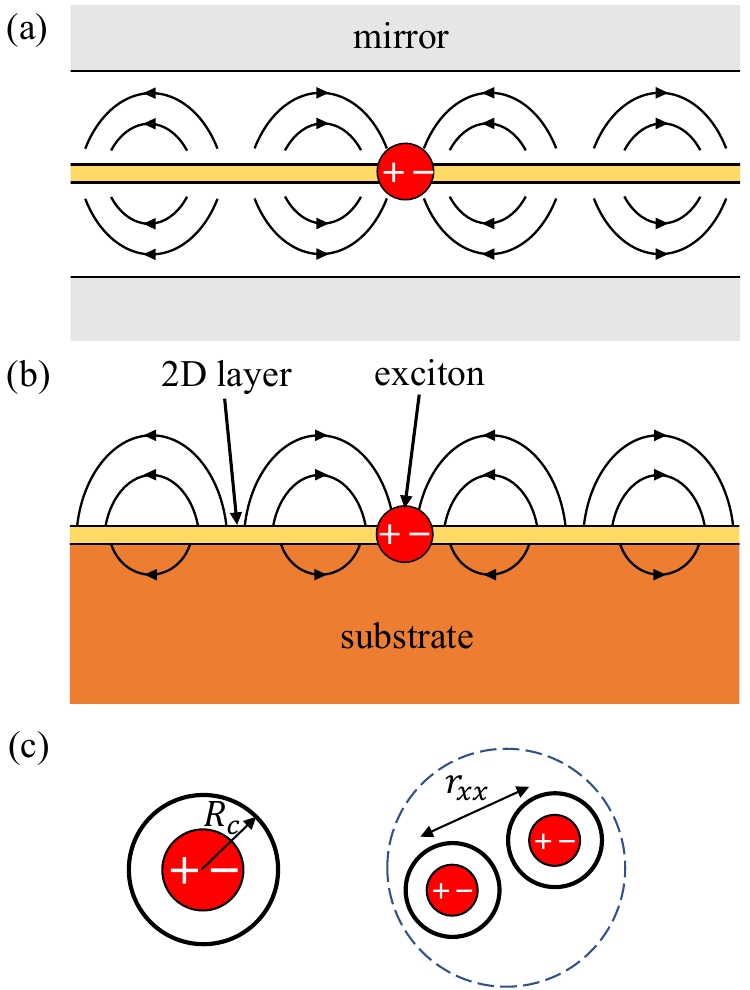}
\end{center}
\caption{Schematic diagrams of the photon cavity (a) and phonon/plasmon cavity (b).
The lines indicate the electric field direction.
(c) Schematic of the heavy-exciton polariton (left) and bipolariton (right)
The ring surrounding the exciton indicates the characteristic 
polariton radius $R_c=(1/m_x\Omega)^{1/2}$, 
with $m_x$ the exciton mass and $\Omega$ the polariton Rabi frequency.
Bipolaritons are well defined when the biexciton binding energy 
$E_{xx}$ is much smaller than $\Omega$, or equivalently
$R_c$ is much smaller than the length scale $r_{xx}=(1/m_xE_{xx})^{1/2}$,
which is of order the bipolariton radius.}
\label{fig:cavities}
\end{figure}

Polaritons in 2D materials, such as phonon polaritons in hexagonal boron nitride
and exciton-polaritons in tranitional metal dichalcogenide (TMD) monolayers,
are a growing research field which provides a controllable platform 
to study polaritons and their interactions \cite{Basov2016, Low2017}.
Motivated by this, we consider a
novel type of exciton-polariton formed by
strong coupling of excitons in a 2D semiconductor 
with surface optical phonons or surface plasmons, see Fig.~\ref{fig:cavities}(a),(b).
In contrast to previous investigations of resonant exciton-phonon coupling \cite{Toyozawa1968, Hermanson1970, Levinson1973},
we consider the regime where the phonon frequency is resonant with transitions across the band gap,
not between internal states of the exciton.
This is similar to the case of plexcitons,
which are hybrid quasiparticles resulting from
resonant coupling between excitons and plasmons in 
metallic nanostructures~\cite{Goncalves2018, Baranov2017, Sun2021}.
By analogy to heavy fermions, we use a common term heavy exciton-polaritons 
for these quasiparticles to emphasize that they have a large effective mass
inherited from a nearly flat dispersion of phonons or plasmons.
Here we analyze the Feshbach scattering resonance between heavy polaritons
and show that they can form bipolariton states near the resonance.
In contrast to previously studied exciton-polaritons in photonic cavities,
where paired states are essentially biexcitons,
heavy polaritons retain their hybrid nature in the bound state [Fig.~\ref{fig:cavities}(c)]
assuming that Rabi frequency exceeds the biexciton binding energy.
We subsequently consider a system with finite densities of polaritons and bipolaritons 
that can exhibit Bose condensation and superfluidity.
We study possible phases of such a condensate, the polariton superfluid (PSF) and bipolariton superfluid (BSF)
~\cite{Radzihovsky2004, Radzihovsky2008, Romans2004, Marchetti2014},
and the associated excitation spectra within a mean-field approximation.

The rest of this paper is organized as follows.
In Sec.~\ref{sec:model} we first discuss single-particle states and compare 
heavy exciton-polaritons to photon exciton-polaritons.
We then include exciton-exciton interactions
and show how bipolariton states emerge near the Feshbach resonance.
In the last part of Sec.~\ref{sec:model}, we study the consequences of heavy bipolariton formation in many-body condensates.
In Sec.~\ref{sec:signatures}, we compute the absorption and luminescence spectra
of the polariton systems and show that they are dominated by collective excitations of the condensate.
We discuss additional experimentally testable predictions of our theory in Sec.~\ref{sec:discussion}.
Details of our calculations are provided in the Appendix.

\section{Model} \label{sec:model}

\subsection{Single-particle states} \label{sec:single-particle}

We consider a simple model of the heavy polariton that captures the essential physics. 
The Hamiltonian is
\begin{align}
H_0 =& \sum_{\bm{k}}\Bigg[\omega_{x,\bm{k}}b_{\bm{k}}^\dag b_{\bm{k}} + \omega_{c,\bm{k}}a_{\bm{k}}^\dag a_{\bm{k}}
+ \frac{\Omega_{\bm{k}}}{2}\left( b_{\bm{k}}^{\dag}a_{\bm{k}}+\mathrm{h.c.} \right) \Bigg], \nonumber \\
\label{eq:hamiltonian1}
\end{align}
which is analogous to what is commonly used to model exciton-polaritons in photonic cavities \cite{Kavokin2011}.
In that context the operators $a_{\bm{k}}^\dag$ and $b_{\bm{k}}^\dag$ respectively create cavity mode and exciton states with momentum $\bm{k}$.
The exciton kinetic energy is $\omega_{x,\bm{k}}=k^2/2m_x$, where $m_x$ is the exciton mass.
In our case $a_{\bm{k}}^\dag$ creates a phonon or a plasmon (Fig.~\ref{fig:cavities})
with energy $\omega_{c,\bm{k}}=\delta$ 
independent of $\bm{k}$ since the dispersion of a surface optical phonon 
(plasmon) is nearly flat compared to that of the exciton.
We refer to parameter $\delta$ as the cavity detuning.
The strength of the coupling is characterized by the Rabi frequency $\Omega_{\bm{k}}$,
i.e., the rate of energy transfer between the two modes,
which we take to be momentum independent for simplicity: $\Omega_{\bm{k}}=\Omega$.
We also ignore the polarization degree of freedom of the exciton.
We use units such that $\hbar=1$ throughout.
\begin{figure}[t]
\begin{center}
\includegraphics[width=3.3in]{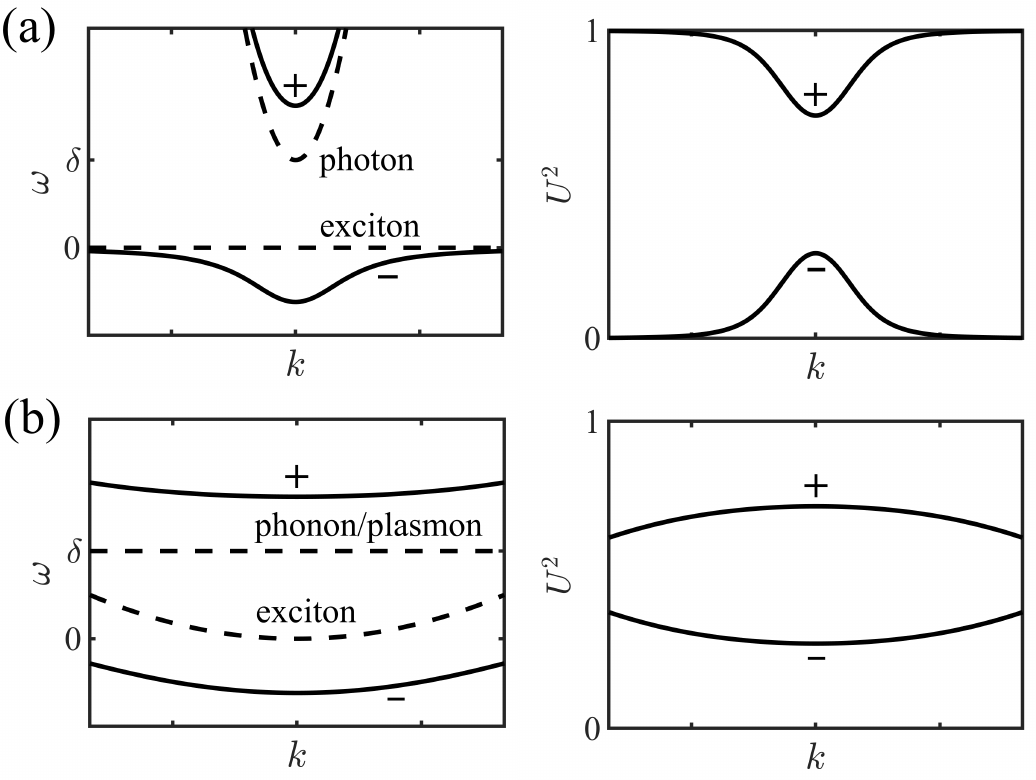}
\end{center}
\caption{Plots of the polariton energy (left) and the exciton fraction (right) versus momentum 
for (a) photon exciton-polaritons and (b) heavy exciton-polaritons.}
\label{fig:polariton_disp}
\end{figure}
The spectrum consists of upper and lower polariton branches with energies
\begin{equation}
\omega_{\pm,\bm{k}} = \frac{1}{2}\left[ \omega_{c,\bm{k}}+\omega_{x,\bm{k}} \pm \sqrt{(\omega_{x,\bm{k}}-\omega_{c,\bm{k}})^2+\Omega_{\bm{k}}^2} \right], 
\label{eq:pol_disp}
\end{equation}
and the exciton fraction in each branch is given by the squares of the Hopfield coefficients 
\begin{equation}
U_{\pm,\bm{k}}^2 = \frac{1}{2}\left[ 1 \pm (\omega_{x,\bm{k}}-\omega_{c,\bm{k}})\big{/}\sqrt{(\omega_{x,\bm{k}}-\omega_{c,\bm{k}})^2+\Omega_{\bm{k}}^2} \right].
\label{eq:hopfield}
\end{equation}

In Fig.~\ref{fig:polariton_disp}, we compare the 
momentum dependence of the polariton energy and Hopfield coefficients 
for conventional photon exciton-polaritons and heavy exciton-polaritons.
Cavity photons have a very steep energy-momentum dispersion which takes the form $\omega_{p,\bm{k}}\simeq\delta+k^2/2m_p$,
with an effective mass $m_p\sim10^{-4}m_x$.
Therefore, for a photonic cavity, the polaritons have a strong dispersion at small $k$, 
but with increasing $k$ outside the light-cone
the lower polariton rapidly becomes mostly exciton and the upper polariton becomes mostly photon.
This is contrasted with the heavy polaritons which have an effective mass of order the 
exciton mass $m_x$ and retain their hybrid character at much larger $k$.

\subsection{Two-particle states}

The interaction between two excitons, viewed as rigid structureless bosons, takes the form
\begin{equation}
H_\mathrm{int} = \frac{1}{2A}\sum_{\bm{k}\bm{k'}\bm{q}} W_{\bm{q}}\, b_{\bm{k}+\bm{q}}^\dag b_{\bm{k'-q}}^\dag b_{\bm{k'}} b_{\bm{k}},
\end{equation}
where $A$ is the area of the system and $W_{\bm{q}}$ is an attractive interaction that supports the biexciton bound state.
The total Hamiltonian is $H=H_0+H_\mathrm{int}$, with $H_0$ defined in Eq.~\eqref{eq:hamiltonian1}.
We may estimate the bipolariton binding energy $E$ from the well-known formula 
for weak attractive interactions in 2D \cite{Landau2013}:
\begin{equation}
E \simeq E_\Lambda \exp(-\frac{2\pi}{|V_p|m_r}),
\label{eq:be}
\end{equation}
with $V_p<0$ the polariton interaction strength at zero momentum, $m_r$ the reduced mass of the two polaritons forming the bound state,
and $E_\Lambda$ a high-energy cutoff determined by the range of the interaction. 
For photon exciton-polaritons, $m_r\sim m_p\sim10^{-4}m_x$, so the binding energy is exponentially small,
as mentioned earlier in Sec.~\ref{sec:introduction}
(see also Appendix~\ref{scattering}).
In contrast, for heavy polaritons $m_r$ and $V_p$ are of the same order as those of the exciton.
Therefore, strongly bound bipolaritons are expected near the biexciton resonance.
\begin{figure}[t]
\begin{center}
\includegraphics[width=3.2in]{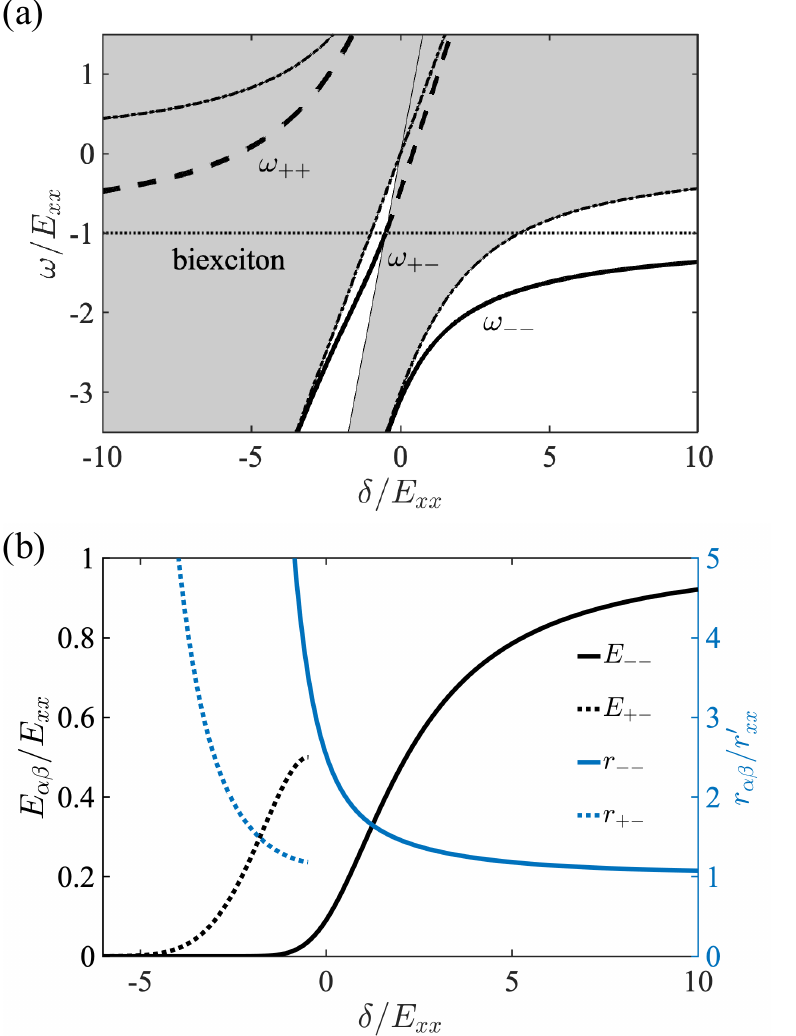}
\end{center}
\caption{(a) Energies of the bipolariton states versus detuning $\delta$ for $\Omega=3E_{xx}$. 
The solid and dashed black lines are the energies of lower-lower, upper-lower, and upper-upper bipolaritons,
denoted respectively by $\omega_{--}$, $\omega_{+-}$, and $\omega_{++}$.
The horizontal dotted line is the biexciton energy, the negative of the biexciton binding energy $E_{xx}$.
The dashed-dotted lines are the sum of the energies of two polaritons at zero momentum, 
with the two-polariton continuum shown by the shaded regions above.
(b) The binding energies $E_{--}=2\omega_--\omega_{--}$ and $E_{+-}=\omega_++\omega_--\omega_{+-}$ 
and corresponding radii of the bipolariton states versus detuning $\delta$.
The radii $r_{ss'}$ are defined by $r_{ss'}^2=\int d^2r\,r^2|\psi_{ss'}(\bm{r})|^2$, with $\psi_{ss'}(\bm{r})$
the bipolariton wavefunction, and $r_{xx}'=(2/3m_xE_{xx})^{1/2}$ is the biexciton radius for the contact potential model.}
\label{fig:bipolariton}
\end{figure}
We find the bipolariton energies from the poles of the exciton two-particle scattering matrix.
To simplify the calculation we choose $W_{\bm{q}}$ to be momentum independent, i.e., a contact interaction,
and subsequently eliminate it in favor of the biexciton binding energy $E_{xx}$ to obtain the renormalized scattering matrix $\mathcal{T}(\bm{K},\omega)$,
which depends on the total incoming momentum $\bm{K}$ and energy $\omega$ of the particles.
We give an explicit formula for the scattering matrix, along with details of the calculation, in Appendix~\ref{scattering}.

Here and in the following calculations we choose $\Omega=3E_{xx}$ for the Rabi frequency.
In Fig.~\ref{fig:bipolariton}(a) we plot the bipolariton energies versus cavity detuning $\delta$.
There are two Feshbach resonances. The first one occurs when the energy of a biexciton (the horizontal dotted line)
conincides with the energy of two lower polaritons (the lower boundary of the continuum), 
$2\omega_-=-E_{xx}$, all at momenta $\bm{k}=0$.
The second resonance is found where the biexciton energy 
is equal to the sum of the energies of a lower and upper polariton, 
$\omega_+ + \omega_- = -E_{xx}$.
In both cases there are bipolariton states lying below the continuum of unbound two-particle states,
e.g., $\omega = \omega_-(\bm{k}) + \omega_-(-\bm{k})$.
In principle, two upper polaritons can also form a quasi-bound state
but it lies in the continuum and is damped.
Figure~\ref{fig:bipolariton}(b) shows the binding energies and radii of the states outside the continuum versus $\delta$. 
The bipolariton states become more loosely bound with decreasing $\delta$ as the exciton fraction $u_-^2$ decreases,
e.g., for two lower polaritons $m_r\propto u_-^{-2}$ and $V_p\propto u_-^4$,
hence their binding energy decreases towards zero in accordance with Eq.~\eqref{eq:be} while their radius diverges.

\subsection{Phase diagram and excitations}

Many-body physics of a system with a finite concentration of polaritons 
may be described by a two-channel effective Hamiltonian \cite{Marchetti2014}
that explicitly includes polariton ($\xi_1$) and bipolariton ($\xi_2$) fields:
\begin{align}
H_\mathrm{eff} =& \sum_{i=1,2}\sum_{\bm{k}} (\omega_{i,\bm{k}}-\mu_i)\xi_{i,\bm{k}}^\dag \xi_{i,\bm{k}} \nonumber \\
&+ \frac{1}{2A}\sum_{ij}\sum_{\bm{k}\bm{k'}\bm{q}} g^{ij}_{\bm{k}\bm{k'}\bm{q}}\, \xi_{i,\bm{k}+\bm{q}}^\dag \xi_{j,\bm{k'}-\bm{q}}^\dag \xi_{j,\bm{k'}} \xi_{i,\bm{k}} \nonumber \\
&+ \frac{1}{\sqrt{2A}}\sum_{\bm{k}\bm{q}} \left( \alpha_{\bm{k}\bm{q}} \xi_{1,\bm{q}/2+\bm{k}}^\dag \xi_{1,\bm{q}/2-\bm{k}}^\dag \xi_{2,\bm{q}} + \mathrm{h.c.} \right).
\label{eq:H_eff}
\end{align}
We assume that only the ``$-$'' polaritons, which are the lower energy states, are present.
Parameter $\mu$ is the chemical potential; $\mu_1 = \mu$, $\mu_2 = 2\mu$;
$\omega_{1,\bm{k}}=\omega_{-,\bm{k}}$ and $\omega_{2,\bm{k}}$ [defined by Eq.~\eqref{eqn:T_--} below]
are the energies of the polariton and bipolariton, respectively,
and $\alpha$ is the polariton-bipolariton coupling.
Parameters $g^{ij}$ are repulsive background interactions which depend on the Hopfield coefficents according to
\begin{equation}
g^{ij}_{\bm{k}\bm{k'}\bm{q}}=U_{-,\bm{k}}U_{-,\bm{k}'}U_{-,\bm{k}+\bm{q}}U_{-,\bm{k}'-\bm{q}}\tilde{g}^{ij}/m_x,
\end{equation}
and $\tilde{g}^{ij}$ are dimensionless interaction strengths, which we assume to be constant.
The parameters $\alpha$ and $\omega_{2,\bm{k}}$ are determined by an expansion 
of the polariton-polariton scattering matrix near the bipolariton pole (see Appendix~\ref{scattering}):
\begin{equation}
\begin{split}
\mathcal{T}_{--}(\bm{k},\bm{k'},\bm{K},\omega) 
=& U_{-,\bm{K/2}+\bm{k}} U_{-,\bm{K/2}-\bm{k}} U_{-,\bm{K/2}+\bm{k'}} U_{-,\bm{K/2}-\bm{k'}} \\ &\times\mathcal{T}(\bm{K},\omega)
\simeq \frac{\alpha_{\bm{k}\bm{K}} \alpha_{\bm{k'}\bm{K}}}{\omega-\omega_{2,\bm{K}}}.
\label{eqn:T_--}
\end{split}
\end{equation}
\begin{figure}[t]
\begin{center}
\includegraphics[width=3.5in]{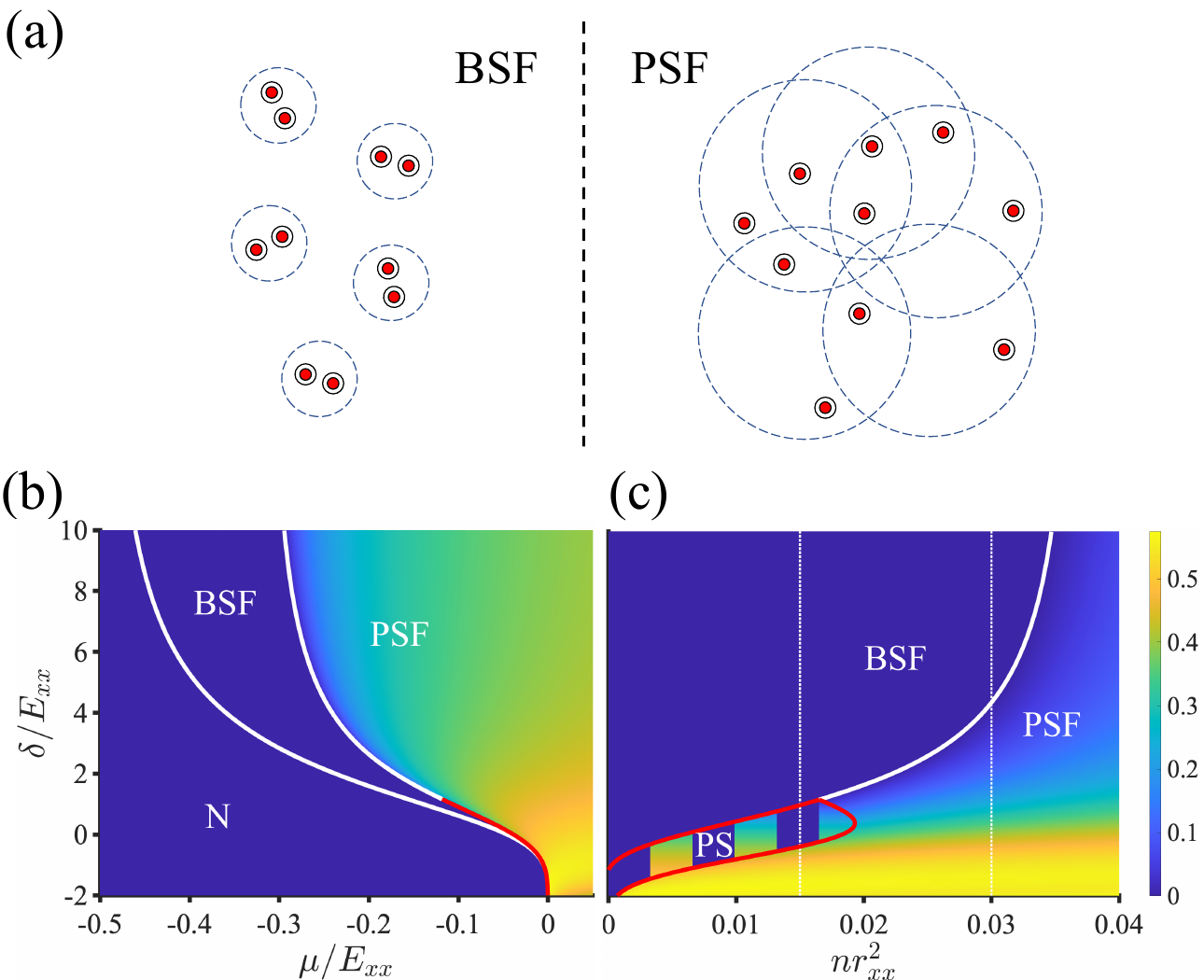}
\end{center}
\caption{(a) Schematic of polariton distribution in the BSF and PSF phases.
(b),(c) Ratio of single polariton condensate density to the total density at zero temperature
as a function of cavity detuning $\delta$ and chemical potential $\mu$ (b) or density $n$ (c). 
The background scattering parameters are $\tilde{g}^{11}=6$, $\tilde{g}^{12}=9$, and $\tilde{g}^{22}=20$.
The length scale $r_{xx}=(1/m_xE_{xx})^{1/2}$ is of order the biexciton radius.
The normal phase is labeled by N, the polariton superfluid by PSF,
the bipolariton superfluid by BSF, and phase separated regions by PS.}
\label{fig:phase_diagram}
\end{figure}

The system exhibits a transition from a polariton to bipolariton superfluid
with changing density and detuning.
In the polariton superfluid phase (PSF) both polaritons and bipolaritons condense; 
in the bipolariton superfluid phase (BSF) only bipolaritons condense.
Following Refs. \citenum{Radzihovsky2008, Marchetti2014}, 
we calculate the mean-field phase diagram at zero temperature using the two-channel model of Eq.~\eqref{eq:H_eff}.
We caution the reader that these phase diagrams are not suitable for immediate comparison 
with experiment because quantum fluctuations beyond mean-field theory
can significantly shift the phase boundaries~\cite{Marchetti2014}.
Additionally, there is a significant uncertainty in the values of
phenomenological parameters $\tilde{g}_{ij}$ (see more below).

We plot the ratio of the single polariton condensate density
to the total density in Fig.~\ref{fig:phase_diagram}(b),(c),
showing the BSF to PSF phase transition with decreasing $\delta$, 
which can be either first-order or continuous.
In the case of a first-order transition, there is a region of phase separation between 
the PSF phase and the BSF or normal phase.
The mean-field phase boundary of the continuous transition is described by the equation 
\begin{equation}
E_{--}=(g_{22}/2-g_{12})n+\alpha\sqrt{2n},
\label{eq:phase_boundary}
\end{equation}
with $E_{--}$ the bipolariton binding energy.
This equation relates the total density $n$ to the cavity detuning $\delta$.
Note that the parameters $\tilde{g}^{ij}$ are difficult to determine even for simplified microscopic models;
furthermore, they depend crucially on details such as the exciton spin structure not included in our simple model.
We crudely estimate these parameters by requiring that the BSF-PSF transition occur at a critical density
$n_M$ in accordance with the Mott criterion $n_Mr_{--}^2\sim0.03$ \cite{Fogler2014, Maezono2013}.
Here $r_{--}$ is the bipolariton radius which is plotted versus detuning in Fig.~\ref{fig:bipolariton}(b).
Associated with this criterion is the picture of particle distribution in real space, Fig.~\ref{fig:phase_diagram}(a).
In the BSF phase, polaritons form a dilute gas of bound pairs.
As the system approaches the phase transition, the binding energy of these pairs decreases
while their radius becomes larger so they begin to overlap in space.
As a result, a significant fraction of these pairs dissociate leading to the formation of a single-polariton condensate
in addition to the bipolariton condensate in the PSF phase.

As discussed below in Sec.~\ref{sec:signatures}, the optical response of the PSF and BSF phases is dominated by their collective excitations.
The excitation spectrum has one gapless acoustic mode and one gapped mode in each phase \cite{Radzihovsky2008}.
In the BSF phase, the gapless mode is due to phase oscillations of the bipolariton condensate
and the gapped mode is due to pair-breaking of bipolaritons into two polaritons.
The energy of the gapped mode is 
\begin{align}
E_1^\mathrm{BSF} = \left\{\left[\tfrac12(E_{--}-g_{22}n_2+2g_{12}n_2)\right]^2-\alpha^2n_2\right\}^{1/2},
\label{eq:gapped_mode_BSF}
\end{align}
where $n_1$ is the single polariton density and $n_2$ is the bipolariton density.
In the PSF phase, the gapless and gapped modes are due to 
in-phase and out-of-phase fluctuations of polariton and bipolariton fields, respectively.
The energy of the gapped mode is
\begin{align}
E_1^\mathrm{PSF} = \Big\{& \left[E_{--}-g_{12}n_1-g_{22}n_2+2g_{11}n_1+2g_{12}n_2-\alpha\sqrt{n_2} \right]^2 \nonumber \\
&+ \alpha n_1\left[\alpha-2g_{12}\sqrt{n_1} + (4g_{11}+g_{22})\sqrt{n_2}\right] \Big\}^{1/2}.
\label{eq:gapped_mode_PSF}
\end{align}
Both $E_1^\mathrm{BSF}$ and $E_1^\mathrm{PSF}$ vanish along the phase boundary given by Eq. ~\eqref{eq:phase_boundary}. 
Far from the phase transition, and neglecting the Feshbach coupling $\alpha$,
the energies of the gapped modes have a simple interpretation.
We may write $E_1^\mathrm{PSF}=\tilde{\omega}_2-2\tilde{\omega}_1$,
where $\tilde{\omega}_1=g_{11}n_1+g_{12}n_2$ and $\tilde{\omega}_2=-E_{--}+g_{12}n_1+g_{22}n_2$
are respectively the polariton and bipolariton energies renormalized by interaction.
This corresponds to the energy to create a bipolariton.
Similarly, $E_1^\mathrm{BSF}=\tfrac12(2\tilde{\omega}_1-\tilde{\omega}_2)$ corresponds to the energy 
to break a bipolariton into its constituent polaritons.

\section{Experimental signatures} \label{sec:signatures}

\begin{figure}[ht]
\begin{center}
\includegraphics[width=3.4in]{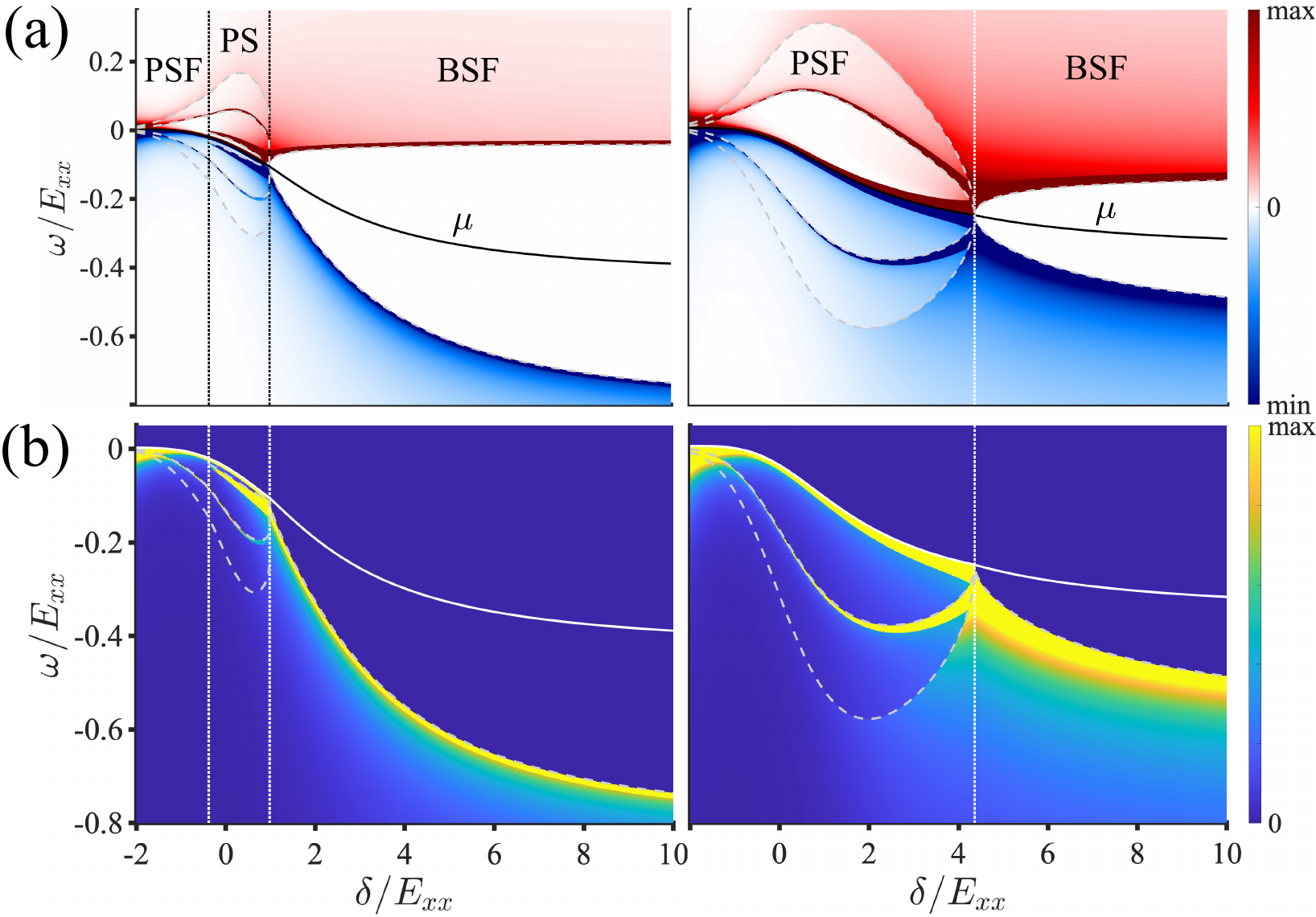}
\end{center}
\caption{Absorption (a) and luminescence (b) spectra at zero temperature 
for density $nr_{xx}^2=0.015$ (left) and $nr_{xx}^2=0.03$ (right).
The chemical potential $\mu$ is denoted by the solid lines, 
and dashed grey lines indicate thresholds associated with the emission of collective excitations.
The absorption spectrum is negative for frequencies below $\mu$, 
corresponding to optical gain, and positive for frequencies above $\mu$.
Since only emission (and not absorption) of collective excitations is possible at zero temperature, 
luminescence only occurs at frequencies below $\mu$.}
\label{fig:spectra}
\end{figure}

One way to distinguish PSF and BSF phases is by measuring the optical absorption or luminescence spectrum. 
The spectrum in both normal and condensed states is determined from 
the collective excitations of the system, which we calculate from the effective Hamiltonian of Eq.~\eqref{eq:H_eff}.
Details of the calculation are given in Appendix~\ref{spectra}.
There are two contributions: direct coupling of polaritons to photons, which gives rise to sharp emission lines,
and coupling of a bipolariton to a photon and polariton, which yields a broad continuum. 
In Fig.~\ref{fig:spectra}, we show plots of the absorption and luminescence spectra 
for a polariton condensate at zero temperature.

The energy of a photon measured in an optical experiment
is the sum (difference) of the chemical potential $\mu$ and excitation energy
corresponding to absorption (emission) of a photon along with emission of a collective excitation.
Therefore, the gapless modes follow the lines $\omega=\mu$
(the solid lines in Fig.~\ref{fig:spectra}),
and the gapped modes are positioned above or below these lines.
Within the mean-field theory, the chemical potential is given by
\begin{equation}
\mu = \begin{cases}
g_{11}n_1 + g_{12}n_2 - \alpha\sqrt{n_1n_2} &\mathrm{ (PSF)} \\
\tfrac12(-E_{--}+g_{22}n_2) &\mathrm{ (BSF)}.
\end{cases}
\end{equation}
The energy $\omega = \mu + E_1^\mathrm{PSF}$ of the collective mode above $\mu$ in the PSF phase
and the energy $\omega = \mu - E_1^\mathrm{BSF}$
below $\mu$ in the BSF phase, can be viewed as the renormalized energy of the bipolariton spectral line.
Far from the transition, this energy is given by the formula
\begin{align}
	\omega = \tilde{\omega}_2 - \tilde{\omega}_1
= -E_{--} + (g_{12} - g_{11})n_1 + (g_{22} - g_{12})n_2\,.
\label{eq:omega_BP}
\end{align}
At large positive $\delta$, the system is in the BSF phase,
where bipolaritons are energetically favored.
The gap between $\mu$ and the absorption or emission threshold
is the energy of the gapped excitation mode,
given by Eq.~\eqref{eq:gapped_mode_BSF}.
The gapless mode, which is due to phase oscillations of the bipolariton condensate,
is not observed in this phase because bipolaritons do not couple directly to light.
With decreasing detuning, the gap decreases and closes at the phase transition,
where single polariton condensation becomes favorable.
In the PSF phase, both gapless and gapped modes are observed.
The energy of the gapped mode is
given by Eq.~\eqref{eq:gapped_mode_PSF}.

We now discuss the relation of our results to previous experiments 
on the polaritonic Feshbach resonance \cite{Takemura2014,Takemura2017, NavadehToupchi2019}. 
Photon exciton-polaritons studied in those experiments 
have a polarization or pseudo-spin degree of freedom
since there are two countercircular polarization states of the photon
that couple to excitons with the same polarization.
Since the interaction between two excitons is attractive (repulsive)
when they have anti-parallel (parallel) spin,
only polaritons with anti-parallel spin can form biexcitons.
In the experiments, the system is pumped with circularly polarized light 
to create a condensate of spin-up polaritons.
A probe beam with the opposite circular polarization then excites
a few spin-down polaritons which interact with the spin-up condensate to form biexcitons.
This results in a shift in the spin-down polariton energy 
observed in the probe transmission spectrum.
The shift changes from positive to negative with decreasing cavity detuning 
as the system is tuned across the Feshbach resonance,
which is often colloquially described as the interaction changing from repulsive to attractive.
The corresponding effect in our model is the change in the collective mode energy
from $-E_1^\mathrm{BSF}$ to $E_1^\mathrm{PSF}$ in Fig.~\ref{fig:spectra}(a)
and crossing between the photon energy $\omega$ in Eq.~\eqref{eq:omega_BP} and chemical potential $\mu$.
In other words, the renormalized bipolariton binding energy
effectively changes from positive in the BSF phase to negative in the PSF phase.
However, the aforementioned experiments have studied a transient state
rather than the true ground state of the polariton system,
since in principle biexcitons should also condense.
\begin{figure}[b]
\begin{center}
\includegraphics[width=3.4in]{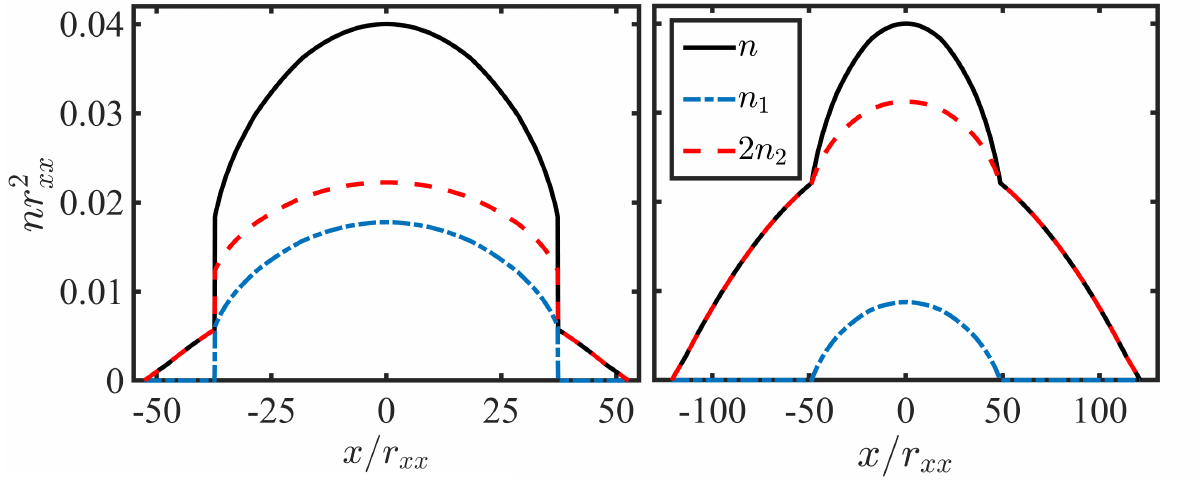}
\end{center}
\caption{Density profiles for polariton condensate in a harmonic trap
for first-order transition (left) and continuous transition (right).
The plots correspond respectively to $\delta/E_{xx}=0$ and $\delta/E_{xx}=2$ in
the phase diagram of Fig.~\ref{fig:phase_diagram}(b).
The trap potential is $V(x)=\frac{1}{2}E_{xx}(x/r_\mathrm{trap})^2$ with $r_\mathrm{trap}=300r_{xx}$.
Here $n_1$ is the single polariton density, $n_2$ is the bipolariton density, and 
$n=n_1+2n_2$ is the total density.} 
\label{fig:density_profile}
\end{figure}

In addition to photoluminescence spectroscopy of a uniform density system, the PSF to BSF transition can be detected
by imaging the polariton condensate confined in a trap.
In Fig.~\ref{fig:density_profile}, we show the density profile in a harmonic trap 
calculated in the Thomas-Fermi approximation by solving $\mu[n(r)]+V(r)=\mu_0$ for $n(r)$,
where $\mu[n]$ is the local chemical potential as a function of density, 
$V(r)=\frac{1}{2}E_{xx}(r/r_\mathrm{trap})^2$ is the trap potential with $r_\mathrm{trap}=300r_{xx}$, 
and $\mu_0$ is the chemical potential at the center of the trap.
Since the local chemical potential $\mu(r)$ decreases with increasing radial coordinate $r$, 
there is a transition from the PSF to BSF phase
which results in a discontinuity in the density $n(r)$ or its derivative $\frac{dn}{dr}$
for a first-order or continuous transition, respectively.
Since $\mu$ is continuous across the phase boundary of the continuous transition, $\Delta\mu=0$, we may relate
the discontinuity in $\frac{\partial \mu}{\partial n}$ to that of $\frac{\partial \mu}{\partial\delta}$ by
$\Delta\frac{\partial\mu}{\partial\delta} = -\frac{dn}{d\delta}\Delta\frac{\partial\mu}{\partial n}$,
with $\frac{dn}{d\delta}$ determined from the phase boundary in Fig.~\ref{fig:phase_diagram}(c).
From Fig.~\ref{fig:spectra}, $\frac{\partial\mu}{\partial\delta}\big\vert_{c-}=-0.022$ and $\frac{\partial\mu}{\partial\delta}\big\vert_{c+}=-0.026$,
where $c-$ ($c+$) indicates that the derivative is evaluated just below (above) the transition point.
This gives $\Delta\frac{\partial\mu}{\partial\delta}\approx -0.004$, 
and since $\frac{dn}{d\delta}\approx 0.002/(E_{xx}r_{xx}^2)$, we have $\Delta\frac{\partial\mu}{\partial n}\approx 2E_{xx}r_{xx}^2$.
Therefore, $\Delta\big(\frac{dn}{dr}\big)^{-1}=\frac{r_\mathrm{trap}^2}{E_{xx}r}\Delta\frac{\partial\mu}{\partial n}\approx 10r_\mathrm{trap}r_{xx}^2$,
which is consistent with the slopes in Fig.~\ref{fig:density_profile}: $\frac{dn}{dr}\big\vert_{c-}\approx-0.5/(r_\mathrm{trap}r_{xx}^2)$ and
$\frac{dn}{dr}\big\vert_{c+}\approx-0.05/(r_\mathrm{trap}r_{xx}^2)$.
This explains the large kink seen in the density profile of Fig.~\ref{fig:density_profile}(right) despite
the small discontinuity of $\frac{\partial\mu}{\partial\delta}$ in Fig.~\ref{fig:spectra}.

\section{Discussion and Outlook} \label{sec:discussion}

We have investigated a novel type of quasiparticle, heavy polaritons, formed by strong coupling
of excitons in a 2D semiconductor with surface optical phonons or plasmons.
Similar to systems of cold atoms, where this phenomenon was first studied,
heavy polaritons exhibit a Feshbach resonance when
the biexciton energy becomes resonant with that of two polaritons.
Such resonances have been recently probed in experiments with exciton-polaritons in photonic cavities.
However, in those experiments the polaritons did not form true bound states.
In contrast, we predict that heavy polaritons bind into bipolaritons near the Feshbach resonance.
For a system with a finite density of polaritons and bipolaritons
we have analyzed the possible phases, polariton (PSF)
and bipolariton (BSF) superfluids, and their collective excitations within a mean field approximation.
We have computed the absorption and luminescence spectra in these phases.

Another observable signature of polariton condensation is formation of quantized vortices.
Such vortices can be induced in the system by external perturbations 
or non-equilibrium flow and subsequently detected by optical imaging \cite{High2012,Leonard2018,Leonard2021}.
According to Ref. \citenum{Radzihovsky2008}, in the PSF phase the Feshbach coupling 
of polaritons and bipolaritons induces a splitting of a $2\pi$ polariton vortex
into two $\pi$ vortices connected by a domain wall.
Observation of this vortex splitting
would confirm the existence of the PSF phase.

In conclusion, we briefly mention possible materials realizations of our results.
Since phonons have frequencies in the terahertz region, 
resonantly coupling them to excitons requires a narrow band gap semiconductor. 
One candidate is gapped bilayer graphene, 
where tunable excitons have been observed with energies around 100 meV \cite{Park2010, Ju2017}.
These excitons can be tuned into resonance with the 
low-loss hyperbolic phonon modes in hBN \cite{Caldwell2014, Giles2017}.
Regarding plasmons, several experiments have demonstrated strong coupling of excitons
in TMD monolayers with plasmons in metallic nanostructures \cite{Sun2021, Qin2020}.
TMDs host strongly bound biexcitons with $E_{xx}\approx50$ meV \cite{You2015}.
Also, a highly controllable realization of 
plasmonic strong local coupling with excitons can be achieved using a nano-optical antenna \cite{Gross2018, Park2019}.
Possible extensions of our work include studying a Josephson-like effect where the optical spectrum in the PSF phase
depends on the relative phase of polariton and bipolariton condensates
and analyzing the collective modes of the polariton condensate in a trap.
A potential application of the heavy polariton Feshbach resonance is the generation of entangled pairs of polaritons \cite{Oka2008},
which is another interesting subject for future study.

\begin{acknowledgements}
We thank L. Radzihovsky, D.N. Basov, and X. Xu for useful discussions.
\end{acknowledgements}

\appendix

\begin{widetext} 

\section{Scattering matrix calculation} \label{scattering}

The scattering matrix satisfies the Bethe-Salpeter equation,
which with the notation $k=(\bm{k},i\omega_n)$ takes the form \cite{BastarracheaMagnani2019}
\begin{equation}
\mathcal{T}(k,k';k+q,k'-q) = W(k,k',q) + \sum_{q'} W(k,k'q')G_x(k+q')G_x(k'-q')\mathcal{T}(k+q',k'-q';k+q;k'-q).
\label{eq:scattering}
\end{equation}
Here $W(k,k',q)$ is the interaction vertex and $G_x(k)$ is the exciton Green's function.
If $W(k,k',q)=W_0$ is frequency and momentum independent, 
then $\mathcal{T}$ depends only on the total incoming momentum and energy $(\bm{K},\omega)$
and Eq.~\eqref{eq:scattering} is solved immediately:
\begin{equation}
\mathcal{T}(\bm{K},\omega)  = \left[ W_0^{-1} - \Pi(\bm{K},\omega) \right]^{-1},
\end{equation}
with the two-exciton propagator (at zero temperature)
\begin{equation}
\Pi(\bm{K},\omega) = \sum_{\bm{q}}\int\frac{d\omega'}{2\pi i} G_x(\bm{K-q},\omega-\omega')G_x(\bm{q},\omega').
\end{equation}
Consider first the case of free excitons, where the Green's function is $G_x^0(\bm{k},\omega) = (\omega-\omega_{x,\bm{k}})^{-1}$, and the propagator is 
\begin{equation}
\Pi_0(\bm{K},\omega) = \sum_{\bm{q}}\int\frac{d\omega'}{2\pi i} G_x^0(\bm{K-q},\omega-\omega')G_x^0(\bm{q},\omega') = \sum_{\bm{q}}\frac{1}{\omega-\omega_{x,\bm{K-q}}-\omega_{x,\bm{q}}}.
\end{equation}
Since the sum diverges logarithmically at large $\bm{q}$, we must impose a momentum cutoff $\Lambda$, 
and we have $\Pi_0(\bm{K}=0,\omega) = -\frac{m_x}{4\pi}\ln(-E_\Lambda/\omega)$, with $E_\Lambda=\Lambda^2/2m_x$.
We eliminate $W_0$ by requiring that the biexciton binding energy $E_{xx}$ is the pole of the of the zero-momentum scattering matrix,
so $W_0^{-1} = -\frac{m_x}{4\pi}\ln(E_\Lambda/E_{xx})$, and the renormalized free exciton scattering matrix takes the form
\begin{equation}
\mathcal{T}_0(\bm{K}=0,\omega) = \frac{4\pi/m_x}{\ln\left(-E_{xx}/\omega \right)},
\end{equation}
which is the universal result for low-energy scattering in 2D \cite{Bolle1984}.
We now consider the case of exciton-polaritons, where the exciton Green's function is 
\begin{equation}
G_x(\bm{k},\omega) = \sum_{s=\pm} \frac{U_{s,\bm{k}}^2}{\omega-\omega_{s,\bm{k}}}.
\end{equation}
The polariton energies $\omega_{\pm,\bm{k}}$ and Hopfield coefficients $U_{\pm,\bm{k}}$ 
are given respectively in Eqs.~\eqref{eq:pol_disp} and \eqref{eq:hopfield} of the main text.
The two-exciton propagator is
\begin{equation}
\Pi(\bm{K},\omega) = \sum_{\bm{q}}\sum_{s,s'=\pm}\frac{U_{s,\bm{K-q}}^2U_{s',\bm{q}}^2}{\omega-\omega_{s,\bm{K-q}}-\omega_{s',\bm{q}}}.
\label{eq:prop}
\end{equation}

For photon exciton-polaritons, the cavity mode dispersion takes the form $\omega_{p,\bm{k}}=\delta+k^2/2m_p$, with $m_p\ll m_x$.
We expand $\Pi(\bm{K}=0,\omega)$ in $m_p/m_x\ll 1$ and find
\begin{align}
\mathcal{T}(\bm{K}=0,\omega) =\frac{4\pi}{m_x}\Bigg\{&\ln(-E_{xx}/\omega) + \frac{m_p}{m_x}\frac{\Omega^2}{\omega^2}\bigg[ -1 + \ln(-(m_x/2m_p)\omega/E_{xx}) 
- \frac{1}{\omega^2+\Omega^2}\big( \omega^2\ln((\omega_++\omega_--\omega)/E_{xx}) \nonumber \\ 
&+ \Omega^2(-\ln(-2\omega/E_{xx})+\ln((2\omega_+-\omega)/E_{xx})+\ln((2\omega_--\omega)/E_{xx})) \big) \bigg] + \ldots \Bigg\}^{-1}.
\label{eq:photon_bipolariton}
\end{align}
Here $\omega_\pm = \frac{1}{2}\left( \delta\pm\sqrt{\delta^2+\Omega^2} \right)$ are the polariton energies at zero momentum.
This shows that $\mathcal{T}(\bm{K}=0,\omega)$ has a pole at the biexciton binding energy which is slightly shifted from $E_{xx}$ and 
acquires a small imaginary part in the two-polariton continuum.
In addition, there are bipolariton poles at energies below $\omega_s+\omega_{s'}$, but it is apparent from Eq.~\eqref{eq:photon_bipolariton} 
that the bipolariton binding energy and spectral weight are exponentially small in the large mass ratio $m_x/m_p$, 
in agreement with Eq.~\eqref{eq:be} of the main text.

For heavy polaritons we take $\omega_{c,\bm{k}}=\delta$ constant. 
Then the integral in Eq.~\eqref{eq:prop} may be evaluated analytically at $\bm{K}=0$ and we find
\begin{equation}
\mathcal{T}(\bm{K}=0,\omega) = \frac{-(4\pi/m_x)\left[(2\delta-\omega)^2+\Omega^2\right]}{\Omega^2\ln(2(\omega_++\omega_--\omega)/E_{xx}) 
+ (2\delta-\omega)^2 \left[-\ln((2\delta-\omega)/E_{xx})+\ln((2\omega_+-\omega)/E_{xx})+\ln((2\omega_--\omega)/E_{xx}) \right]},
\end{equation}
The bipolariton energies $\omega_{ss',\bm{K}}$ are poles of $\mathcal{T}(\bm{K},\omega)$.
The scattering matrix between two polaritons $s$ and $s'$ with incoming momenta $\bm{K/2}+\bm{k}$ and $\bm{K/2-k}$
and outgoing momenta $\bm{K/2}+\bm{k'}$ and $\bm{K/2-k'}$ is given by 
\begin{equation}
\mathcal{T}_{ss'}(\bm{k},\bm{k'},\bm{K},\omega) = U_{s,\bm{K/2}+\bm{k}}U_{s',\bm{K/2-k}}U_{s,\bm{K/2}+\bm{k'}}U_{s',\bm{K/2-k'}}\mathcal{T}(\bm{K},\omega).
\end{equation}
The polariton Green's functions are $G_s(\bm{k},\omega) = (\omega-\omega_{s,\bm{k}})^{-1}$ and we define
$\tilde{G}_{ss'}(\bm{k},\bm{K},\omega) = \int\frac{d\omega'}{2\pi i} G_s(\omega-\omega',\bm{K/2-k})G_{s'}(\omega',\bm{K/2}+\bm{k})$, or
\begin{equation}
\tilde{G}_{ss'}(\bm{k},\bm{K},\omega) = \frac{1}{\omega-\omega_{s,\bm{K/2-k}}-\omega_{s',\bm{K/2}+\bm{k}}}.
\end{equation}
In the spectral vicinity of the bipolariton resonance, we have \cite{Ivanov1998}
\begin{equation}
\tilde{G}_{ss'}(\bm{k},\bm{K},\omega)\mathcal{T}_{ss'}(\bm{k},\bm{k'},\bm{K},\omega)\tilde{G}_{ss'}(\bm{k'},\bm{K},\omega) \simeq \frac{\psi_{ss'}(\bm{k},\bm{K})\psi_{ss'}(\bm{k'},\bm{K})}{\omega-\omega_{ss',\bm{K}}},
\label{eq:pole_exp}
\end{equation}
which defines the bipolariton wavefunction $\psi_{ss'}(\bm{k},\bm{K})$.
In our approximation
\begin{equation}
\psi_{ss'}(\bm{k},\bm{K}) = \mathcal{N}_{\bm{K}}\frac{U_{s,\bm{K/2-k}}U_{s',\bm{K/2}+\bm{k}}}{\omega_{ss',\bm{K}}-\omega_{s,\bm{K/2-k}}-\omega_{s',\bm{K/2}+\bm{k}}},
\end{equation}
with $\mathcal{N}_{\bm{K}}$ a normalization factor such that $\sum_{\bm{k}}|\psi_{ss'}(\bm{k},\bm{K})|^2=1$.
The bipolariton wavefunction is used to compute the optical spectra, see Appendix~\ref{spectra}.

\section{Calculation of absorption and luminescence spectra} \label{spectra}

Two processes contribute to the absorption and luminescence spectra: direct coupling of polaritons to photons 
and coupling of a bipolariton to a photon and polariton \cite{Haug1984, Hanamura1977}. 
The relevant diagrams for calculating photon self-energy are shown in Fig.~\ref{fig:diagrams2}.
The filled circle denotes the photon-polariton coupling vertex.
In terms of the photon-exciton coupling $\mu_x$,
it is given by
\begin{equation}
\mu_{1,\bm{q}}=U_{-,\bm{q}}\mu_x,
\end{equation}
where $U_{-,\bm{q}}$ is the Hopfield coefficient.
The open circle denotes the photon-polariton-bipolariton coupling vertex, 
given by
\begin{equation}
\mu_{12,\bm{k},\bm{q}}=U_{-,\bm{k}}U_{-,\bm{q}+\bm{k}}^2\psi((\bm{q}-\bm{k})/2,\bm{q}+\bm{k})\mu_x,
\end{equation}
with $\psi$ the bipolariton wavefunction defined by Eq.~\eqref{eq:pole_exp}.

\subsubsection{Normal state}

In the normal state, only the first two diagrams of Fig.~\ref{fig:diagrams2} contribute. 
The Matsubara Green's functions are
\begin{equation}
G_i(\bm{k},i\omega_n) = \frac{1}{i\omega_n-\tilde{\omega}_{i,\bm{k}}},
\end{equation}
where $i=1,2$ for the polariton and bipolariton respectively.
Here $\tilde{\omega}_{1,\bm{k}}$ and $\tilde{\omega}_{2,\bm{k}}$ are the polariton and bipolariton energies 
shifted by interaction (see Sec.~\ref{sec:model}).
The photon self-energy is
\begin{equation}
\Pi(\bm{q},i\omega_n) = |\mu_{1,\bm{q}}|^2G_1(\bm{q},i\omega_n)-\sum_{\bm{k}}\sum_{i\omega_n'}|\mu_{12,\bm{k},\bm{q}}|^2G_1(\bm{k},i\omega_n')G_2(\bm{q}+\bm{k},i\omega_n+i\omega_n').
\end{equation}
Evaluating the sum over $i\omega_n'$, we obtain the absorption spectrum:
\begin{equation}
A(\bm{q},\omega) = -2\Im\Pi(\bm{q},\omega-\mu) = |\mu_{1,\bm{q}}|^22\pi\delta(\omega-\mu-\tilde{\omega}_{1,\bm{q}}) + \sum_{\bm{k}}|\mu_{12,\bm{k},\bm{q}}|^2[n_B(\tilde{\omega}_{1,\bm{k}})-n_B(\tilde{\omega}_{2,\bm{q}+\bm{k}})]2\pi\delta(\omega-\mu+\tilde{\omega}_{1,\bm{k}}-\tilde{\omega}_{2,\bm{q}+\bm{k}}).
\end{equation}
with 
\begin{equation}
n_B(\omega)=\big(e^{\beta\omega}-1\big)^{-1}
\end{equation}
the Bose distribution function, $\beta$ being the inverse temperature.
The luminescence spectrum is \cite{Haug1984}
\begin{align}
I(\bm{q},\omega)=n_B(\omega-\mu)A(\bm{q},\omega) =& |\mu_{1,\bm{q}}|^2n_B(\tilde{\omega}_{1,\bm{q}})2\pi\delta(\omega-\mu-\tilde{\omega}_{1,\bm{q}}) \nonumber \\
&+ \sum_{\bm{k}}|\mu_{12,\bm{k},\bm{q}}|^2[1+n_B(\tilde{\omega}_{1,\bm{k}})]n_B(\tilde{\omega}_{2,\bm{q}+\bm{k}})2\pi\delta(\omega-\mu+\tilde{\omega}_{1,\bm{k}}-\tilde{\omega}_{2,\bm{q}+\bm{k}}).
\end{align}
For $\bm{q}=0$ and far from the Feshbach resonance, this spectrum consists of a sharp peak at $\omega=\mu+g_{11}n_1+g_{12}n_2$ 
and a continuum below the energy $\omega = \mu - E_{--} + (g_{11}-g_{12})n_1+(g_{12}-g_{22})n_2$, similar to Eq.~\eqref{eq:omega_BP}.
In principle, this spectrum is qualitatively different from what is shown in Fig.~\ref{fig:spectra}(b),
and so it can be used to identify the condensation transition.

\begin{figure}[h]
\begin{center}
\includegraphics[width=3.2in]{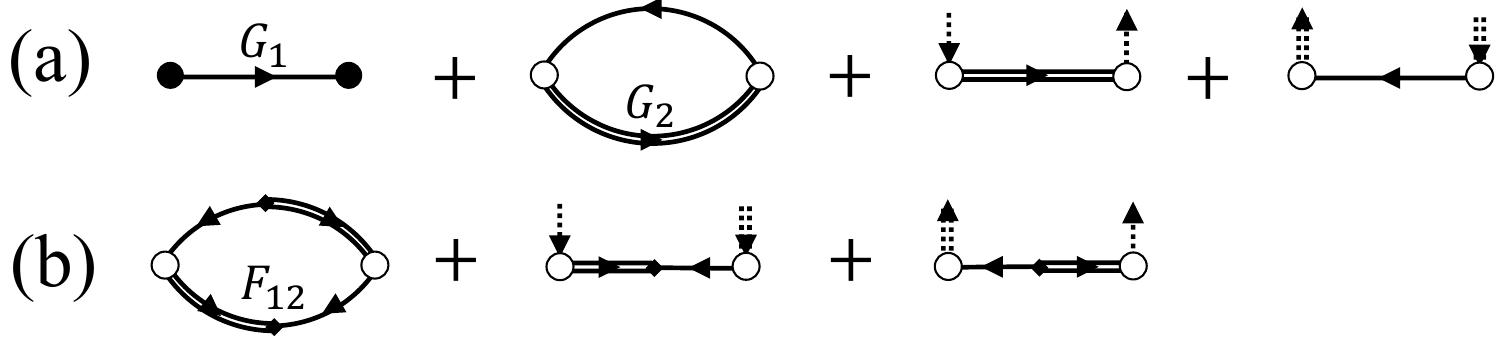}
\end{center}
\caption{Diagrams for calculating photon self-energy including (a) normal and (b) anomalous contributions.
In the normal state only the first two diagrams in (a) are non-vanishing.
Single and double lines denote exciton and biexciton Green's functions, respectively.
The filled circle denotes the photon-polariton coupling vertex, and the open circle
is the photon-polariton-bipolariton coupling vertex.
Dotted lines indicate particles going into or out of the condensate.}
\label{fig:diagrams2}
\end{figure}

\subsubsection{Condensed state}

In the condensed state, there are two modes: a gapless mode $E_{0,\bm{k}}$ and gapped mode $E_{1,\bm{k}}$.
The Matsubara Green's functions may be written
\begin{equation}
G_i(\bm{k},i\omega_n) = \sum_{\sigma=0,1}\sum_{s=\pm}\frac{s\big(u_{i,\bm{k}}^{\sigma,s}\big)^2}{i\omega_n-sE_{\bm{k}}^\sigma}.
\end{equation}
Here the $u_{i,\bm{k}}^{\sigma,s}$ are Bogolubov coefficients that satisfy $\sum_{\sigma,s}s\big(u_{i,\bm{k}}^{\sigma,s}\big)^2=1$.
We also need the anomalous Green's functions 
\begin{equation}
F_{ij}(\bm{k},i\omega_n) = -\sum_{\sigma=0,1}\sum_{s=\pm}\frac{su_{i,\bm{k}}^{\sigma,s}u_{j,\bm{k}}^{\sigma,-s}}{i\omega_n-sE_{\bm{k}}^\sigma}.
\end{equation}
Expressions for the energies and Bogolubov coefficients in terms of 
the parameters in the effective Hamiltonian of Eq.~\eqref{eq:H_eff} are given in Ref. \citenum{Radzihovsky2008}.
The photon self-energy is
\begin{align}
\Pi(\bm{q},i\omega_n) =& |\mu_{1,\bm{q}}|^2G_1(\bm{q},i\omega_n)-\sum_{\bm{k}}\sum_{i\omega_n'}|\mu_{12,\bm{k},\bm{q}}|^2G_1(\bm{k},i\omega_n')G_2(\bm{q}+\bm{k},i\omega_n+i\omega_n') \nonumber \\
&- \sum_{\bm{k}}\sum_{i\omega_n'}|\mu_{12,\bm{k},\bm{q}}|^2F_{12}(\bm{k},i\omega_n')F_{21}^*(\bm{q}+\bm{k},i\omega_n+i\omega_n') \nonumber \\
&+ |\mu_{12,\bm{0},\bm{q}}|^2\left[ n_1G_2(\bm{q},i\omega_n) + n_2G_1(\bm{q},-i\omega_n) - 2\sqrt{n_1n_2}F_{12}(\bm{q},i\omega_n) \right].
\end{align}
We find for the absorption spectrum
\begin{align}
A(\bm{q},\omega) =& -2\Im\Pi(\bm{q},\omega-\mu) = \sum_{n=1}^3 A_n(\bm{q},\omega), \\ 
A_1(\bm{q},\omega) =& -|\mu_{1,\bm{q}}|^2\sum_{\sigma,s}s\big(u_{1,\bm{q}}^{\sigma,-s}\big)^22\pi\delta\big(\omega-\mu+sE_{\bm{q}}^\sigma\big), \nonumber \\
A_2(\bm{q},\omega) =& \sum_{\sigma,\sigma'}\sum_{s,s'}\sum_{\bm{k}}|\mu_{12,\bm{k},\bm{q}}|^2 
ss'\left[ \big(u_{1,\bm{k}}^{\sigma,s}\big)^2\big(u_{2,\bm{q}+\bm{k}}^{\sigma',s'}\big)^2 + u_{1,\bm{k}}^{\sigma,s}u_{2,\bm{k}}^{\sigma,-s}u_{1,\bm{q}+\bm{k}}^{\sigma',-s'}u_{2,\bm{q}+\bm{k}}^{\sigma',s'} \right]
\big[n_B\big(sE_{\bm{k}}^\sigma\big)-n_B\big(s'E_{\bm{q}+\bm{k}}^{\sigma'}\big)\big] \nonumber \\
&\times2\pi\delta\big(\omega-\mu+sE_{\bm{k}}^\sigma-s'E_{\bm{q}+\bm{k}}^{\sigma'}\big), \nonumber \\
A_3(\bm{q},\omega) =& -|\mu_{12,\bm{0},\bm{q}}|^2\sum_{\sigma,s}s\left[ n_1\big(u_{2,\bm{q}}^{\sigma,-s}\big)^2+n_2\big(u_{1,\bm{q}}^{\sigma,s}\big)^2-2\sqrt{n_1n_2}u_{1,\bm{q}}^{\sigma,s}u_{2,\bm{q}}^{\sigma,-s} \right]2\pi\delta\big(\omega-\mu+sE_{\bm{q}}^\sigma\big). \nonumber
\end{align}
The luminescence spectrum is
\begin{align}
I(\bm{q},\omega) =& I_0(\bm{q},\omega) + n_B(\omega-\mu)A(\bm{q},\omega) = \sum_{n=0}^3I_n(\bm{q},\omega), \\
I_0(\bm{q},\omega) =& \left( |\mu_{1,\bm{0}}|^2n_1 + |\mu_{12,\bm{0},\bm{0}}|^2n_1n_2 \right)(2\pi)^3\delta(\omega-\mu)\delta(\bm{q}), \nonumber \\
I_1(\bm{q},\omega) =& |\mu_{1,\bm{q}}|^2\sum_{\sigma,s}s\big(u_{1,\bm{q}}^{\sigma,-s}\big)^2\big[1+n_B\big(sE_{\bm{q}}^\sigma\big)\big]2\pi\delta\big(\omega-\mu+sE_{\bm{q}}^\sigma\big), \nonumber \\
I_2(\bm{q},\omega) =& \sum_{\sigma,\sigma'}\sum_{s,s'}\sum_{\bm{k}}|\mu_{12,\bm{k},\bm{q}}|^2ss'\left[ \big(u_{1,\bm{k}}^{\sigma,s}\big)^2\big(u_{2,\bm{q}+\bm{k}}^{\sigma',-s'}\big)^2
+ u_{1,\bm{k}}^{\sigma,s}u_{2,\bm{k}}^{\sigma,-s}u_{1,\bm{q}+\bm{k}}^{\sigma',s'}u_{2,\bm{q}+\bm{k}}^{\sigma',-s'} \right]\big[1+n_B\big(sE_{\bm{k}}^\sigma\big)\big]\big[1+n_B\big(s'E_{\bm{q}+\bm{k}}^{i\sigma}\big)\big] \nonumber \\
& \times2\pi\delta\big(\omega-\mu+sE_{\bm{k}}^\sigma+s'E_{\bm{q}+\bm{k}}^{\sigma'}\big), \nonumber \\
I_3(\bm{q},\omega) =& |\mu_{12,\bm{0},\bm{q}}|^2\sum_{\sigma,s}s\left[ n_1\big(u_{2,\bm{q}}^{\sigma,-s}\big)^2+n_2\big(u_{1,\bm{q}}^{\sigma,s}\big)^2-2\sqrt{n_1n_2}u_{1,\bm{q}}^{\sigma,s}u_{2,\bm{q}}^{\sigma,-s} \right]\big[1+n_B\big(sE_{\bm{q}}^\sigma\big)\big]2\pi\delta\big(\omega-\mu+sE_{\bm{q}}^\sigma\big). \nonumber
\end{align}
The first term is due to coherent spontaneous emission of photons from the condensate,
and is not present in the absorption spectrum.
Actually, when $q\rightarrow0$ the weight of the peak at $\omega=\mu$ diverges due to the divergence of the Bogoliubov coefficients.
We deal with this by integrating over a finite range of $\bm{q}$, $q<q_c$.
Physically, the cutoff $q_c$ represents a typical momentum due to scattering by inhomogeneities, phonons, etc.,
or the inverse trap size in the case of confined polaritons (Sec.~\ref{sec:signatures}).

\end{widetext}

\bibliographystyle{apsrev4-2}	
\bibliography{citations}

\end{document}